\documentclass[aps,prb,superscriptaddress,twocolumn,showpacs,amsmath,amssymb]{revtex4-2}
\usepackage{graphicx,color}
\usepackage{comment}
\usepackage{mathptmx}
\usepackage{helvet}
\usepackage{bm}
\usepackage{sidecap}%
\usepackage{here}
\newcounter{num}

\newcommand{\Tc}{$T_{\mathrm{c}}$}
\newcommand{\Ts}{$T_{\mathrm{s}}$}
\newcommand{\Tn}{$T_{\mathrm{N}}$}
\newcommand{\Tz}{$T_{\mathrm{0}}$}

\newcommand{\BFA}{$\mathrm{BaFe_{2}As_{2}}$}
\newcommand{\BFAP}{$\mathrm{BaFe_{2}}(\mathrm{As}_{1-x}\mathrm{P}_{x}\mathrm{)_{2}}$}
\newcommand{\BFCA}{$\mathrm{Ba(Fe}_{1-x}\mathrm{Co}_{x}\mathrm{)_{2}As_{2}}$}
\newcommand{\NFCA}{$\mathrm{NaFe}_{1-x}\mathrm{Co}_x\mathrm{As}$}

\newcommand{\cm}{$\mathrm{cm^{-1}}$}
\newcommand{\Ag}{$A_{\mathrm{1g}}$}
\newcommand{\Bg}{$B_{\mathrm{1g}}$}
\newcommand{\Bgg}{$B_{\mathrm{2g}}$}
\newcommand{\cma}{$\mathrm{cm}^{-1}$}

\newcommand{\dwave}{$d_{x^2-y^2}$}
\newcommand{\BaFeCoAs}{$\mathrm{Ba(Fe}_{1-x}\mathrm{Co}_{x}\mathrm{)_{2}As_{2}}$}
\newcommand{\chinem}{$\chi^{B_{1g}}_0$}

\begin{document}

\title{Superconducting gap and nematic resonance at the quantum critical point observed by Raman scattering in \BFAP}

\author{T. Adachi}
\email[]{}
\affiliation{Department of Physics, Graduate School of Science, Osaka University, 
Toyonaka, Osaka 560-0043, Japan}

\author{M. Nakajima}
\affiliation{Department of Physics, Graduate School of Science, Osaka University, 
Toyonaka, Osaka 560-0043, Japan}

\author{Y. Gallais}
\affiliation{Laboratoire Mat\'{e}riaux et Ph\'{e}nom\`{e}nes Quantiques (UMR 7162 CNRS), Universit\'{e} de Paris, Bat. Condorcet, 75205 Paris Cedex 13, France}

\author{S. Miyasaka}
\affiliation{Department of Physics, Graduate School of Science, Osaka University, 
Toyonaka, Osaka 560-0043, Japan}

\author{S. Tajima}
\affiliation{Department of Physics, Graduate School of Science, Osaka University, 
Toyonaka, Osaka 560-0043, Japan}

\begin{abstract}
We report comprehensive temperature and doping-dependences of the Raman scattering spectra for \BFAP\ ($x =$ 0, 0.07, 0.24, 0.32, and 0.38), focusing on the nematic fluctuation and the superconducting responses. With increasing $x$, the bare nematic transition temperature estimated from the Raman spectra reaches $T =$ 0 K at the optimal doping, which indicates a quantum critical point (QCP) at this composition. In the superconducting compositions, in addition to the pair breaking peaks observed in the \Ag\ and \Bg\ spectra, another strong \Bg\ peak appears below the superconducting transition temperature which is ascribed to the nematic resonance peak. The observation of this peak indicates significant nematic correlations in the superconducting state near the QCP in this compound.

\end{abstract}


\maketitle
\section{Introduction}
More than a decade has passed since the discovery of iron-based superconductors (IBSs) \cite{Kamihara}. However, its superconducting (SC) mechanism is not yet well understood. A smoking gun is the fact that a magnetic ordered phase is adjacent to the SC phase, and thus the spin fluctuation is a strong candidate for the pairing glue. On the other hand, recently nematic fluctuations have also been observed \cite{Chu, Patz, Gallais2, Bohmer1, Lu, Mirri, Bohmer2, Diog, Hosoi, Toyoda}, which have attracted much attention. Several theoretical and experimental studies suggest that nematic fluctuations may play a key role in the superconductivity in IBSs \cite{Sato, Liu, Yamase1, Led, Agatsuma, Led_2, Labat}. However, whether nematic quantum criticality is relevant for the appearance of superconductivity in IBS, or not, remains largely unsettled.

Raman scattering spectroscopy is a powerful tool to investigate the electronic properties in solids. In particular, the symmetry-resolved sensitivity enables us to directly access nematic behavior without any external field such as uniaxial strain. Recently, Raman scattering experiments have been performed on many IBSs such as \BFCA, \NFCA, FeSe, and LaFeAsO \cite{Gallais2, Thor,Massat,Kaneko}. These results indicate that there exist nematic fluctuations in the tetragonal phase of several IBSs. From the doping dependence of Raman scattering spectra in \BFCA\ and \NFCA, a nematic quantum critical point (QCP) has been revealed near the magnetic critical point close to optimal superconducting transition temperature (\Tc) \cite{Gallais2, Thor,Kre}. Moreover, for the superconducting \BFCA, a nematic resonance mode was observed below the SC gap energies near the nematic QCP \cite{Gallais3}. A similar peak seen in \NFCA\ was attributed to an in-gap collective mode in the nematic channel, consistently with the nematic resonance scenario \cite{Thor}. In \BFCA, however, one concern is that SC gap features in the Raman scattering spectra of non-nematic channel are weak \cite{Bohm}, likely due to the disorder effects introduced by Co-substitution for Fe. 

Here we have chosen \BFAP\ system, because it is considered to be a less disordered system than \BFCA\ \cite{Ishida,Nakajima2}, and clear fingerprints of quantum criticality have been observed in thermodynamic measurements \cite{Shishido, Pene, Walmsley}. Therefore, it is an attractive system for the study of nematic fluctuations and related phenomena, but so far no systematic Raman scattering study has been reported on this compound.

Here, we present a systematic study of Raman scattering on \BFAP\ over a wide range of P-compositions ($x$s). At all the studied compositions, the nematic fluctuations are observed above structural transition temperature (\Ts) or \Tc. The bare nematic transition temperature \Tz\ reaches 0 K near the optimal doping, which implies the existence of a nematic QCP at this composition. In the SC state, a clear pair breaking peak is observed in the \Ag\ symmetry for $x =$ 0.32 and 0.38. It is ascribed to the SC gap of the hole pockets. Moreover, another strong peak is observed in the \Bg\ symmetry towards the nematic QCP. This peak is ascribed to the nematic resonance peak, indicating the persistence of significant nematic correlations in the SC state near the QCP.

\section{Methods}
Single crystals of \BFAP ($x =$ 0, 0.07, 0.24, 0.32, and 0.38) were grown by a self-flux method as described elsewhere \cite{Nakajima}. The transition temperatures (\Ts\ and \Tc) were determined by resistivity measurements before Raman scattering measurements. Single crystals were cleaved just before being loaded into a cryostat. 
Raman scattering spectra were obtained in the pseudo-backscattering configuration using an Ar laser line (514.5 nm) and a T64000 Jobin-Yvon triple grating spectrometer equipped with a liquid-nitrogen-cooled charge coupled device (CCD) detector. The gratings have density of 1800 gr/mm. For the measurements below \Tc, the laser power was set to 4 mW, while it was 10 mW for the measurements above \Tc. The laser heating in these conditions was estimated to be 4 K and 6-10 K, respectively.

$A_{\mathrm{1g}} + B_{\mathrm{2g}}$, $B_{\mathrm{1g}}$, and \Bgg\ Raman spectra were measured with the incident $(i)$ and scattered $(s)$ light polarizations of $(i, s) = (x’, x’), (x’, y’),$ and $(x, y)$, respectively. Here, $x$ and $y$ are oriented along the directions of Fe-Fe bonds, while $x'$ and $y'$ are along the diagonals of Fe-Fe bonds. The pure \Ag\ spectrum is obtained by the subtraction of $xy$ spectrum from $x'x'$ one.

\section{Results and Discussion}
\begin{figure}[htpb]
  \centering
  \includegraphics[width=80mm,origin=c,keepaspectratio,clip]{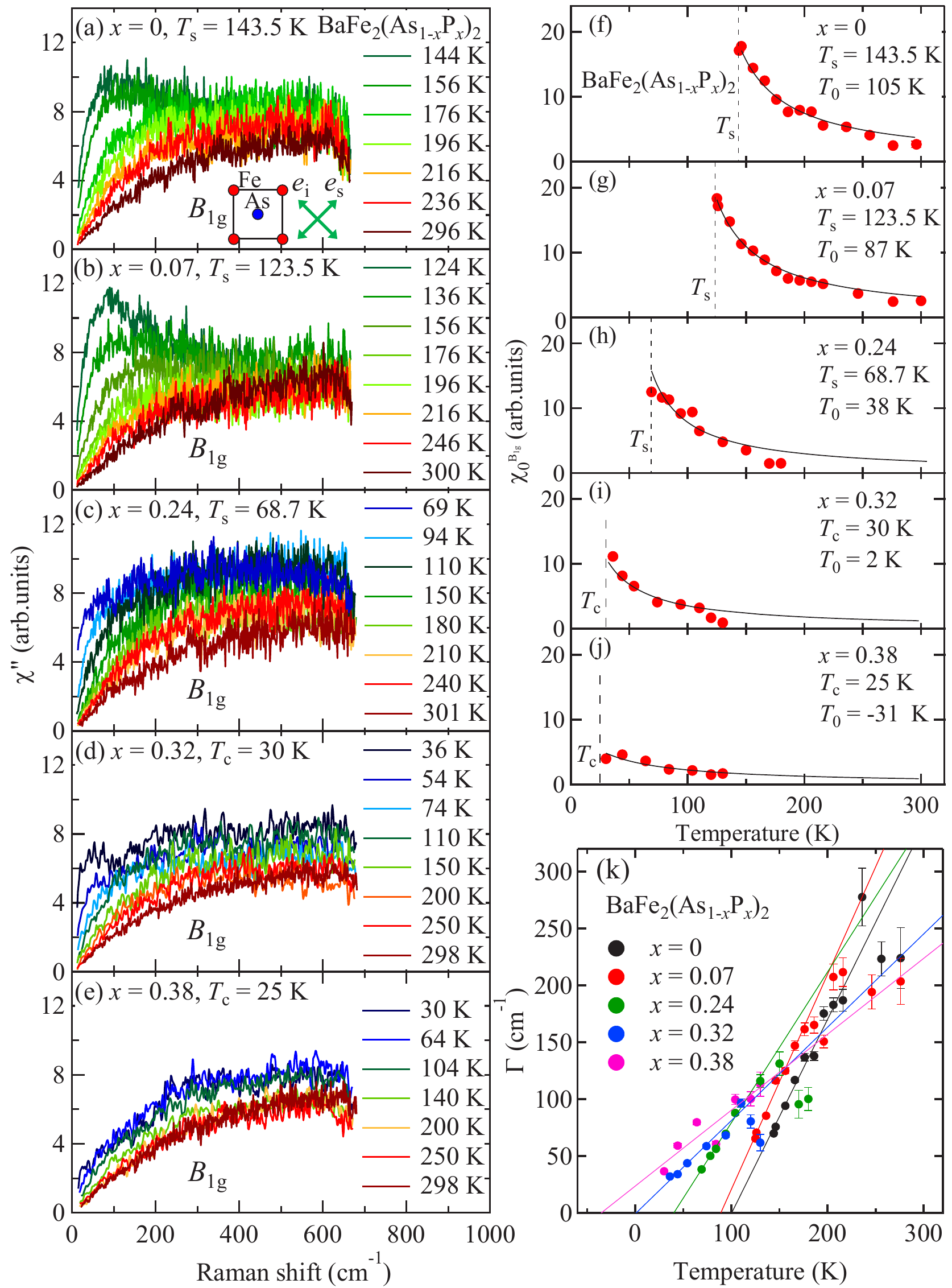}
  \caption{(Color online) (a)-(e) Temperature dependence of \Bg\ Raman scattering spectra for $x =$ 0, 0.07, 0.24, 0.32, and 0.38. (f)-(j) Temperature dependence of the Raman susceptibility, $\chi^{B_{1g}}_{0}$ of \BFAP\ ($x =$ 0, 0.07, 0.24, 0.32, and 0.38). (k) Temperature dependence of the scattering rate $\Gamma$ for $x =$ 0, 0.07, 0.24, 0.32, and 0.38.}
  \label{fig1}
\end{figure}

First, we discuss the nematic fluctuations above \Ts. Figures \ref{fig1}(a)-(e) show the temperature dependence of the \Bg\ Raman responses for \BFAP\ ($x =$ 0, 0.07, 0.24, 0.32, and 0.38). For all the samples, Raman intensity displays a strong enhancement at low energies upon cooling towards \Ts\ or \Tc. This enhancement is suppressed below \Ts\ or \Tc\ and is not observed in the \Ag\ and \Bgg\ symmetries (See Figs. \ref{figS1} and \ref{figS2} in Appendix A and B). A similar behavior has been reported also in some other IBSs and is attributed to enhanced \dwave\ charge fluctuations, namely nematic fluctuations, near $T_s$ \cite{Gallais1}.  The \Bg\ Raman response at low frequencies near \Ts\ is well reproduced using a quasi-elastic peak (QEP) lineshape, $\chi''_{QEP}$, expressed by

\begin{equation}
  \label{QEP}
  \chi''_{QEP}=\frac{A\omega\Gamma}{\omega^2+\Gamma^2},
\end{equation}  
where $A$ is a constant and $\Gamma$ can be interpreted as a quasiparticle scattering rate renormalized by nematic correlations \cite{Gallais1}. In addition to this QEP, a temperature independent $\omega$-linear background was assumed.
The QEP of \Bg\ Raman scattering response is directly connected with the static charge nematic susceptibility, $\chi^{B_{1g}}_0$ via the Kramers-Kronig relation,
\begin{equation}
\label{KK}
\chi^{B_{1g}}_{0}=\frac{2}{\pi}\int_{0}^{\infty} \frac{\chi''_{QEP}}{\omega} d\omega.
\end{equation}
Within a mean-field theory framework, we expect $\chi^{B_{1g}}_{0}$ to follow Curie-Weiss behavior,
\begin{equation}
\label{CWlaw}  
\chi^{B_{1g}}_{0}=\frac{C}{T-T_0},
\end{equation}
where $C$ is a constant and \Tz\ is the charge nematic transition temperature.

Figures \ref{fig1}(f)-(j) present the temperature dependence of $\chi^{B_{1g}}_0$ obtained from the data in Figs. \ref{fig1}(a)-(e) (See Fig. \ref{figS3} in Appendix C). At each P-composition $x$, the data are well fitted to the Curie-Weiss curve as long as the temperature is near \Ts\ ($T \leq T_{\mathrm{s}}$ $+$ 70 K). 

For all the samples, \Tz\ is always lower than \Ts, consistent with previous results on \BaFeCoAs\ \cite{Gallais2}. This is expected because Raman scattering probes the nematic susceptibility in the dynamical limit. The static nematic susceptibility extracted in this limit does not couple to the soft orthorhombic acoustical phonon, and can thus be considered as the bare lattice-free nematic susceptibility \cite{Kontani,Gallais1}. This was confirmed in \BFCA\ by comparing the nematic susceptibility and the shear modulus $Cs$ \cite{Yoshizawa, Gallais2, Kontani}. 
Figure \ref{fig1}(k) shows the temperature dependence of the linewidth ($\Gamma$) of the QEP in Eq. (\ref{QEP}). $\Gamma$ decreases as the temperature approaches \Ts\ or \Tc. In the mean-field theory of Raman scattering near a nematic instability \cite{Gallais1}, this parameter is the quasiparticle scattering rate $\Gamma_0$ renormalized by the nematic correlation length. Assuming a weakly $T$ dependent $\Gamma_0$ near $T_s$,  $1/\Gamma$ should follow a Curie-Weiss temperature dependence near \Ts\ and vanishes at \Tz. Therefore, another estimate of \Tz\ can be deduced by extrapolating the $T$-linear fitting line for $\Gamma$. The obtained \Tz\ values are in broad agreement with those deduced from $\chi^{B_{1g}}_{0}$ in Figs. \ref{fig1}(f)-(j). 

The \Tz\ for \BFA\ in this study is in good agreement with the previous study \cite{Gallais2} and the shear modulus study \cite{Yoshizawa}, whereas it is lower than that obtained by the elastoresistance experiment \cite{Chu}. It has been reported that elastoresistance experiments tend to give higher \Tz\ than Raman scattering and shear modulus ones for reasons which are not clear at present \cite{Bohmer3}.

\begin{figure}[htpb]
  \centering
  \includegraphics[width=85mm,origin=c,keepaspectratio,clip]{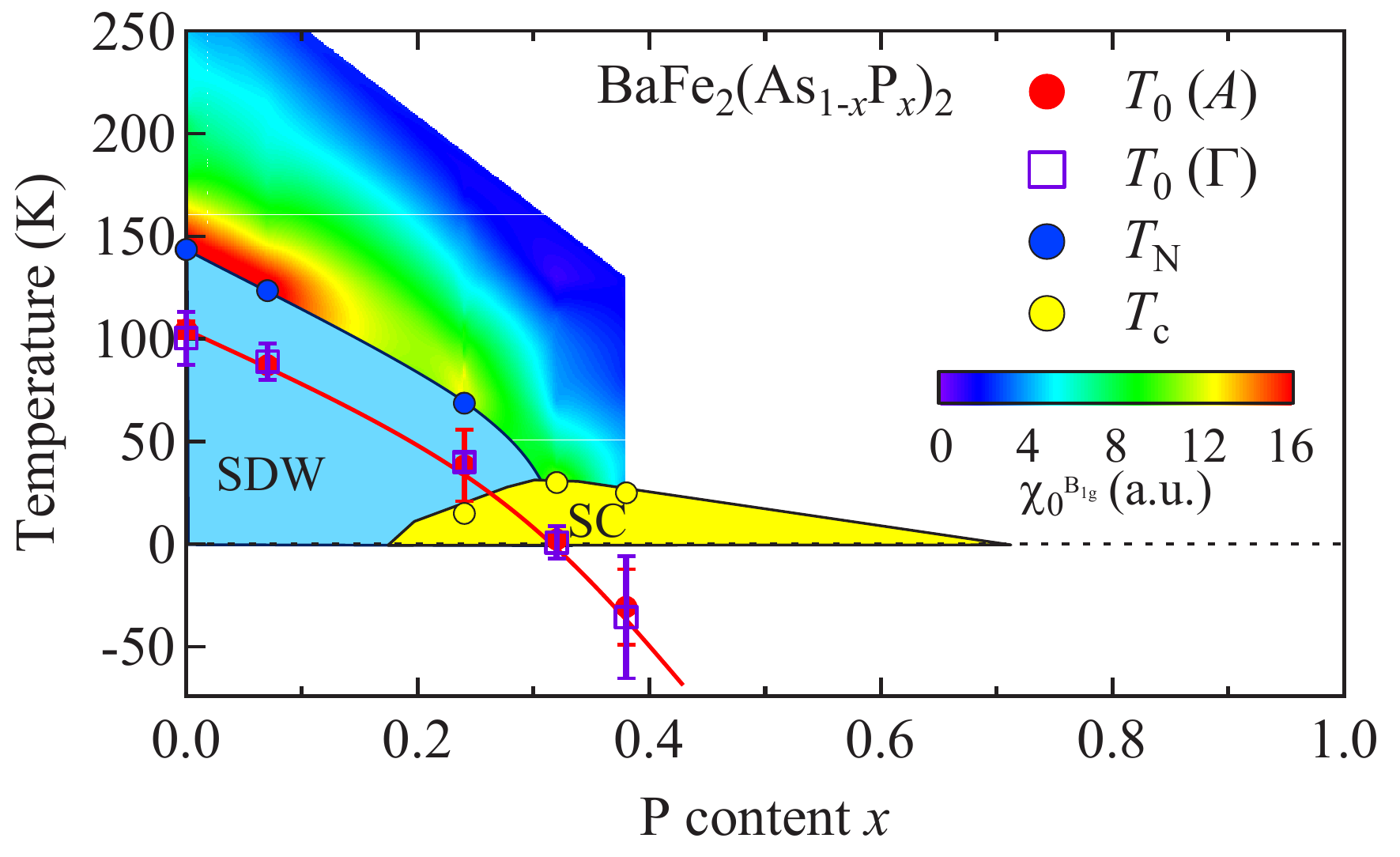}
  \caption{(Color online) Phase diagram of \BFAP. The magnetic phase transition temperature (\Tn) and \Tc\ are taken from Ref. \cite{Nakajima}. The red filled circle indicates \Tz\ derived from the area ($A$) of the QEP, while the purple open square indicates \Tz\ derived from the width ($\Gamma$) of the QEP. The color plot indicates the strength of the \chinem.}
  \label{fig2}
\end{figure}

In Fig. \ref{fig2}, the estimated \Tz\ and $\chi^{B_{1g}}_{0}$ are plotted in the phase diagram of \BFAP\ obtained in the previous studies \cite{Nakajima}. The color plot indicates the strength of the \chinem. \Tz\ reaches 0 K around the optimal doping ($x = 0.32$), indicating a nematic QCP. A similar result was reported by the elastoresistance experiment \cite{Kuo}.

\begin{figure}[htpb]
  \centering
  \includegraphics[keepaspectratio,width=80mm, origin=c]{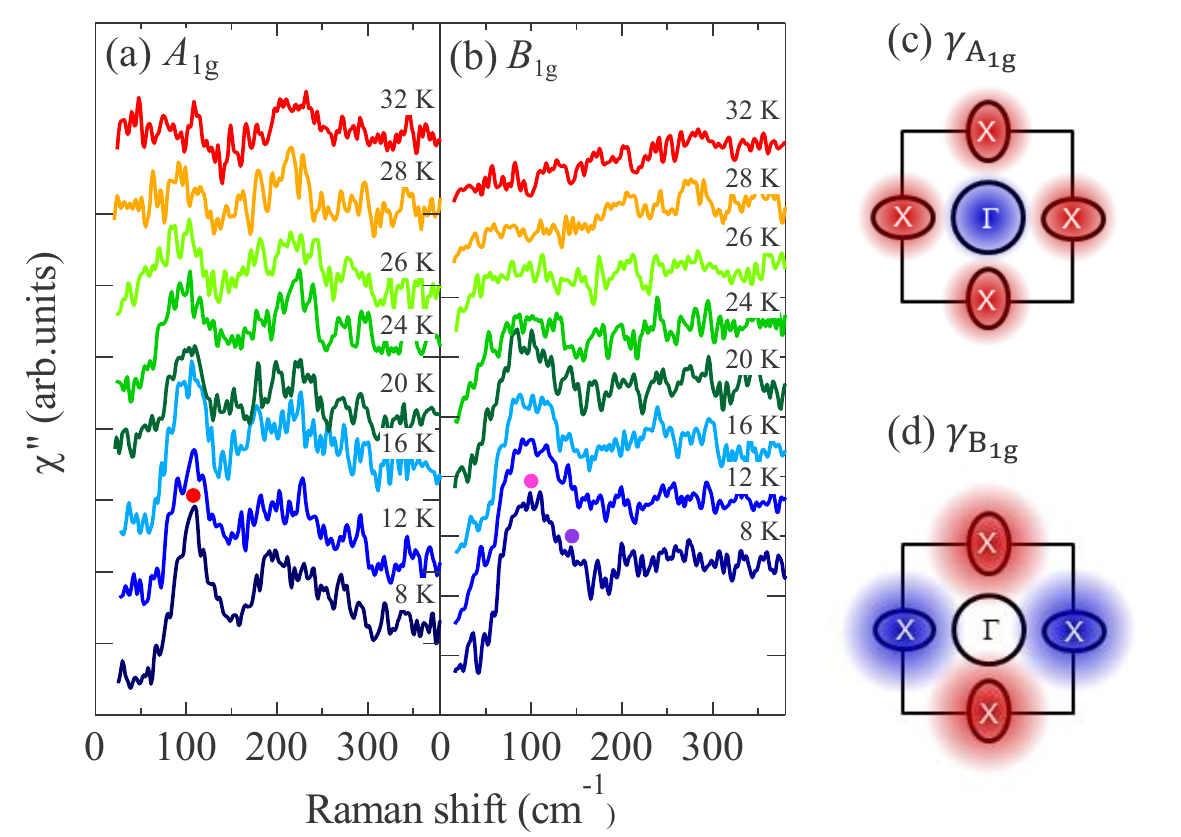}
  \caption{(Color online) (a), (b) Raman scattering spectra in \Ag\ and \Bg\ symmetries of \BFAP\ ($x =$ 0.32, \Tc\ $=$ 30 K) at low temperatures. From the bottom (the lowest temperature) to the top (near \Tc), each spectrum is plotted with each offset. (c), (d) Raman vertices for \Ag\ and \Bg\ symmetries. Red and blue colors indicate positive and negative values, respectively. The Raman vertex for \Ag\ symmetry is illustrated according to the result of the effective-mass approximation \cite{Mazin, Bohm2}. }
  \label{fig3}
\end{figure}

Next, we discuss the Raman responses in the SC state. The temperature dependences of the \Ag\ and \Bg\ Raman scattering spectra below \Tc\ were precisely measured for $x = 0.32$ and 0.38. As for the \Bgg\ Raman response, no spectral difference was observed between above and below \Tc.
As shown in Figs. \ref{fig3}(a) and (b), in both of the \Ag\ and \Bg\ symmetries, distinct features derived from superconductivity are observed. Figure \ref{fig3}(a) demonstrates that a peak develops with decreasing temperature. The \Ag\ peak energy $\Omega_{A_{1g}}$ at 8 K is about $108$ \cma $\sim 13.4$ meV. This energy is comparable to the angle-resolved photoemission spectroscopy (ARPES) data, 2$\Delta_{\mathrm{hole}}\sim 14$ meV where $\Delta_{\mathrm{hole}}$ is the averaged SC gap of hole pockets \cite{Zhang}. Therefore, we assign this peak to the pair-breaking (PB) peak derived from the gap opening on the hole pockets. The \Ag\ spectral shapes near the lowest frequency are almost flat, suggesting a full gap on the hole pockets. Note that the peak around 200 \cm observed in this symmetry exists already in the normal state.

As shown in Fig. \ref{fig3} (b), the spectral shapes in the \Bg\ symmetry are apparently similar to those in the \Ag\ symmetry. In the \Bg\ symmetry, a distinct peak grows around 100 \cma\ at low temperatures. As the peak energy $\Omega^L_{B_{1g}}$ is close to $\Omega_{A_{1g}}$, one may consider that it is a PB peak derived from the hole pockets. However, it is unlikely because the \Bg\ form factor is expected to probe dominantly the electron pockets, as illustrated in Fig. \ref{fig3} (d) \cite{Mazin, Bohm2}. Instead, the weak shoulder structure indicated by the purple mark around 145 \cma ($\sim$ 18 meV) might be a more appropriate candidate of the \Bg\ PB peak, because its energy $\Omega^H_{B_{1g}}$ is close to $2\Delta_{\mathrm{electron}}$ in ARPES \cite{Zhang}. 

\begin{figure}[htpb]
  \centering
  \includegraphics[width=75mm,origin=c,keepaspectratio,clip]{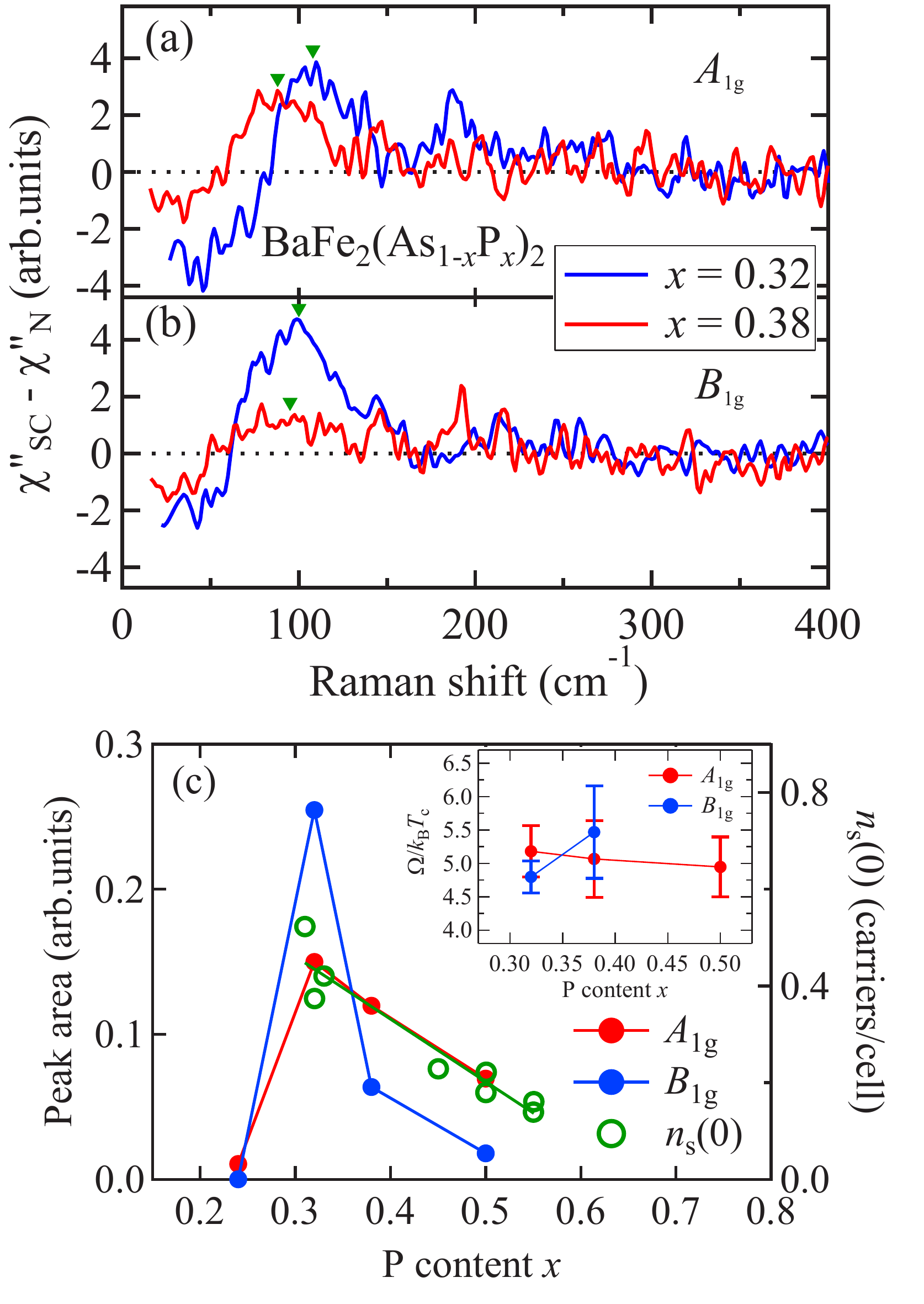}
  \caption{(Color online) (a), (b) Subtracted Raman scattering spectra ($\chi''_{\mathrm{SC}}-\chi''_{\mathrm{N}}$) in \Ag\ and \Bg\ symmetries of \BFAP\ ($x =$ 0.32 and 0.38). The green triangles indicate peak energies of peaks. (c) Doping dependence of the peak area at $\Omega_{A_{1g}}$ and $\Omega^L_{B_{1g}}$ and $n_s(0)$ estimated from specific heat \cite{Diao}. The inset shows the doping dependence of the peak energy ($\Omega$) divided by $k_\mathrm{B}T_\mathrm{c}$ in \Ag\ and \Bg\ symmetries. The data at $x = 0.50$ is taken from Ref. \cite{Wu}.}
  \label{fig4}
\end{figure}

Figures \ref{fig4}(a) and (b) compare the subtracted Raman spectra ($\chi''_{\mathrm{SC}}-\chi''_{\mathrm{N}}$) for $x = 0.32$ and 0.38 in \Ag\ and \Bg\ symmetries, where $\chi''_{\mathrm{SC}}$ and $\chi''_{\mathrm{N}}$ is the Raman susceptibility at 8K and the temperature just above \Tc, respectively.
In the \Ag\ spectra, one can clearly see that $\Omega_{A_{1g}}$ decreases with doping, namely, with decreasing \Tc. While the spectral weight of the \Ag\ peak slightly decreases with doping, that of \Bg\ at $\Omega^L_{B_{1g}}$ is strongly suppressed with increasing $x$.
This tendency of intensity decrease with doping continues to strongly overdoped region \cite{Wu}. At $x =$ 0.5, the PB peak in the \Ag\ symmetry is clearly observed, while the \Bg\ spectrum does not show any distinct peak. 

The doping dependences of the peak intensities in ($\chi''_{\mathrm{SC}}-\chi''_{\mathrm{N}}$) are shown in Fig. \ref{fig4}(c). The data of $x =$ 0.50 \cite{Wu} was normalized with our data at 200 \cma. The PB peak area is proportional to the density of Cooper pairs weighted by the square of the Raman vertex in the BCS framework \cite{Blanc}. In fact, the doping dependence of the \Ag\ peak area follows that of the superfluid density $n_\mathrm{s}(0)$ obtained by the specific heat measurement \cite{Diao}. This is because the specific heat observes mainly heavy quasiparticles which are generally located in the hole bands responsible for \Ag\ PB peak.

The behavior of the \Bg\ peak is, however, very different from that of the \Ag\ peak. The \Bg\ peak intensity of $x = 0.32$ is almost four times larger than that of $x = 0.38$ while \Tc\ of $x = 0.32$ (= 30 K) is only 5 K higher than that of $x = 0.38$ (= 25 K). The enhancement of the \Bg\ peak intensity towards the nematic QCP has also been observed in $\mathrm{NaFe}_{1-x}\mathrm{Co}_x\mathrm{As}$ and \BaFeCoAs\ \cite{Thor, Gallais3, Chau}. Recently, Gallais {\it et al} pointed out that sufficiently close to a nematic QCP the \Bg\ PB peak can be transformed into a nematic resonance mode which arises due to nematic correlations between quasiparticles \cite{Gallais3}. According to this theory, the bare PB peak intensity at $\omega=2\Delta$ gradually decreases close to the QCP and the nematic resonance peak appears at an energy $\Omega_r$ lower than $2\Delta$.

This scenario is in good agreement with the present result in the following senses. (i) The peak energy at $\Omega^L_{B_{1g}}$ is lower than 2$\Delta$ on the electron pockets observed in the present study as well as the ARPES experiment \cite{Zhang}. (ii) The transformation from the PB peak to the nematic resonance also explains naturally the strong enhancement of the \Bg\ peak intensity towards the QCP. (iii) Although with increasing $x$ the \Bg\ peak becomes too weak and broad to discuss its \Tc-scaling, $\Omega^L_{B_{1g}}$ does not seem to scale with \Tc, while $\Omega_{A_{1g}}$ almost does as shown in the inset of Fig. \ref{fig4}(c). It is the same trend with \BFCA\ and consistent with the theory \cite{Gallais3}. (iv) Importantly the nematic quantum critical behavior was observed only for the \Bg\ peak at $\Omega^L_{B_{1g}}$ but not for the other PB peaks, which represents an intrinsic feature of nematicity. 

It should be noted that a nematic resonance is expected to arise irrespective of the microscopic origin of nematicity such as charge (Pomeranchuk instability) \cite{Zhai, Hart} or orbital \cite{Onari, Yamase2} or spin \cite{Fer, Kara,Hino}. Whatever the origin of nematic fluctuations is, the presence of nematic resonance mode within a SC gap and its enhancement towards the QCP strongly suggest a close relation between the nematic fluctuation and superconductivity. An intriguing question is whether these enhanced nematic correlations deep in the SC state can play a role in the divergent-like behavior observed in the London penetration depth of \BFAP\ close to the QCP, near $x =$ 0.3 \cite{Pene}.

\section{Conclusion}
In conclusion, we have systematically investigated the Raman scattering spectra for \BFAP ($x =$ 0, 0.07, 0.24, 0.32, and 0.38). The nematic fluctuations above \Ts\ or \Tc\ were observed at all the studied compositions including the SC ones. The bare nematic transition temperature \Tz\ becomes 0 K near the optimal doping ($x =$ 0.32), indicating the existence of a nematic QCP. In the SC state at $x =$ 0.32 and 0.38, in addition to the pair breaking peaks for the gaps on the hole and electron pockets, we observed a strong \Bg\ peak that is much stronger than the PB peak and correlates with the normal state nematic fluctuations (i.e. QCP behavior). From its doping and symmetry dependent behaviors, this peak can be ascribed to a nematic resonance mode. The present results strongly suggest a firm relation between the superconductivity and the nematic fluctuation in this compound.

\section*{ACKNOWLEDGMENTS}
This work was supported by Grants-in-Aid for Scientific Research from JSPS, Japan. T.A. acknowledges the Grant-in-Aid for JSPS Fellows.

\section*{Appendix A: \Ag\ and \Bgg\ spectra}
\begin{figure}[H]
  \centering
  \includegraphics[width=85mm,origin=c,keepaspectratio,clip]{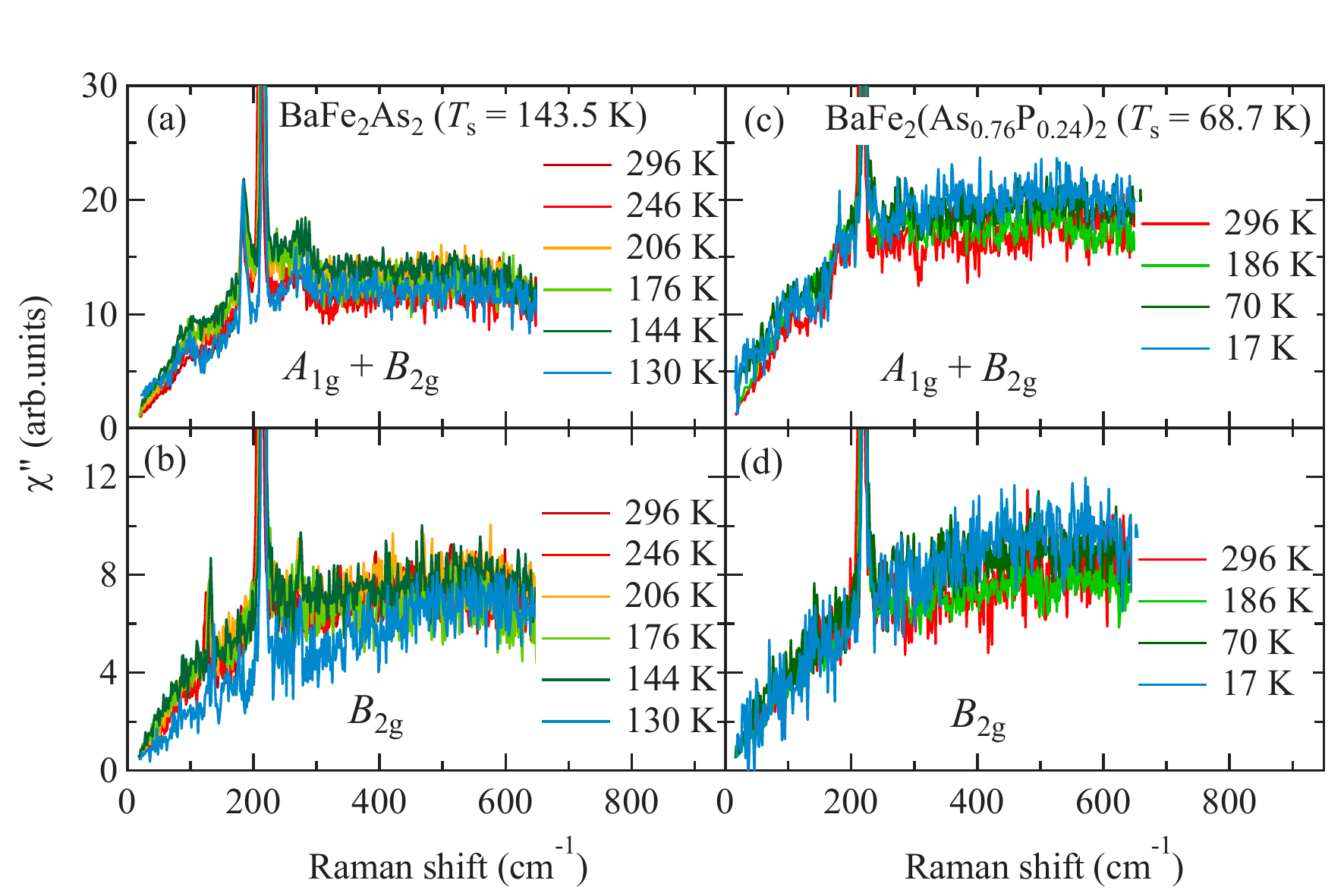}
  \caption{(Color online) Temperature dependence of the $A_{\mathrm{1g}} + B_{\mathrm{2g}}$ and \Bgg\ Raman scattering spectra for \BFAP\ ($x =$ 0 and 0.24). }
  \label{figS1}
\end{figure}

Figure \ref{figS1} shows the temperature dependence of the $A_{\mathrm{1g}} + B_{\mathrm{2g}}$ and \Bgg\ Raman responses for \BFAP\ ($x =$ 0, 0.24). In our measurements, no hump structures could be observed in the \Ag\ and \Bgg\ symmetries above \Ts\ for all the studied compositions. Moreover, there is almost no temperature dependence of the spectra in these symmetries above \Ts.

\section*{Appendix B: \Bg\ spectra around \Ts}
\begin{figure}[H]
  \centering
  \includegraphics[width=80mm,origin=c,keepaspectratio,clip]{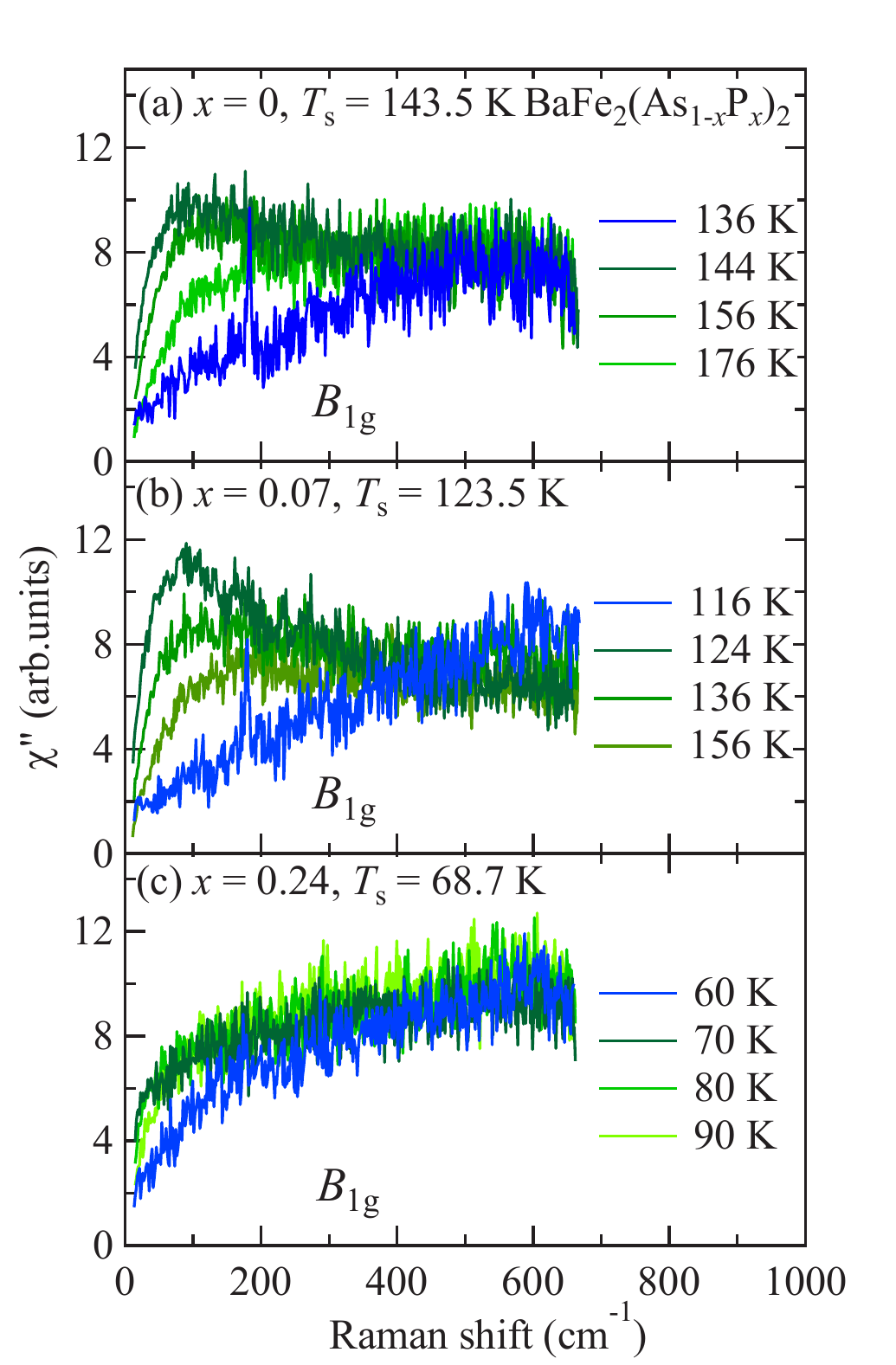}
  \caption{(Color online) Temperature dependence of the \Bg\ Raman scattering spectra for \BFAP\ ($x =$ 0, 0.07, and 0.24). }
  \label{figS2}
\end{figure}

The \Bg\ spectra above and below \Ts\ are presented in Fig. \ref{figS2}. The hump structure rapidly disappears when the sample experiences the structural transition to the low-temperature orthorhombic phase, while it grows with decreasing temperature at $T >$ \Ts. It is noted that the sharp peak around 180 \cma\ below \Ts\ is the $A_g$ phonon mode which is the same as the \Ag\ phonon mode seen in the tetragonal phase. It can be observed in the $x'y'$ polarization configuration only in the orthorhombic phase.

\section*{Appendix C: Determination of \Tz}
\begin{figure}[H]
  \centering
  \includegraphics[width=85mm,origin=c,keepaspectratio,clip]{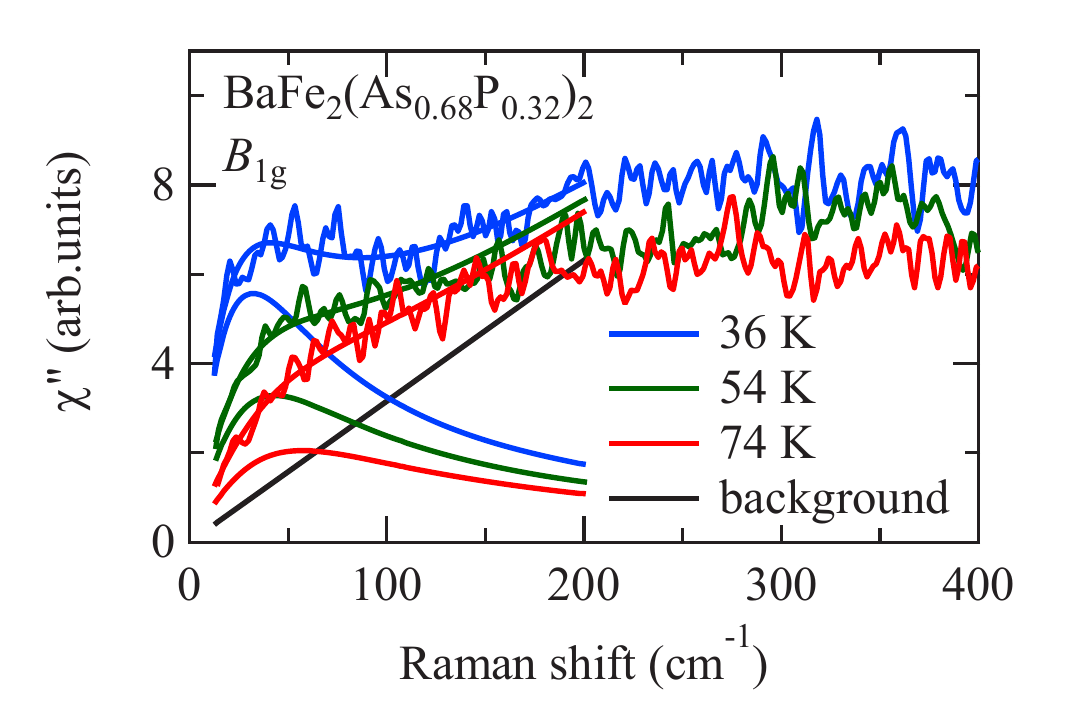}
  \caption{(Color online) Fits of the \Bg\ Raman responses at several temperatures for $\mathrm{BaFe_{2}}(\mathrm{As}_{0.68}\mathrm{P}_{0.32}\mathrm{)_{2}}$. Each fitted line is decomposed into the QEP at each temperature and the background.}
  \label{figS3}
\end{figure}

In our analysis, we assumed that the electronic background in the \Bg\ Raman response is $T$-independent in the same manner as the previous study \cite{Gallais2}. Then, we can decompose the \Bg\ Raman response into the quasi-elastic peak (QEP) and the temperature-independent $\omega$-linear background as long as the temperature is near \Ts\ ($T \leq T_{\mathrm{s}} +$ 70 K) and the frequency is lower than 200 \cm. Figure \ref{figS3} presents \Bg\ Raman responses and fitted lines at several temperatures for $\mathrm{BaFe_{2}}(\mathrm{As}_{0.68}\mathrm{P}_{0.32}\mathrm{)_{2}}$. The \Bg\ Raman responses at low frequencies are well reproduced by this fitting and the nematic susceptibility can be extracted. 

\end{document}